\documentclass[12pt]{article}
\usepackage{graphicx}

\input{epsf}
\usepackage{cite}
\def\@fmsl@sh#1#2#3{\m@th\ooalign{$\hfil#1\mkern#2/\hfil$\crcr$#1#3$}}
 \def\eq#1\en{\begin{equation}#1\end{equation}}
\def\s[#1,#2]{[#1\stackrel{\star}{,}#2]}
\def\sx[#1,#2]{[#1\stackrel{\star_{x}}{,}#2]}

\newcommand{\nc}{\newcommand}
\nc{\beq}{\begin{equation}}
\nc{\eeq}{\end{equation}}
\nc{\beqa}{\begin{eqnarray}}
\nc{\eeqa}{\end{eqnarray}}

\def\gsim{\mathrel{\rlap{\lower4pt\hbox{\hskip1pt$\sim$}}
    \raise1pt\hbox{$>$}}}       %greater than or approx. symbol

\begin{document}
\makeatletter
\def\fmslash{\@ifnextchar[{\fmsl@sh}{\fmsl@sh[0mu]}}
\def\fmsl@sh[#1]#2{%
  \mathchoice
    {\@fmsl@sh\displaystyle{#1}{#2}}%
    {\@fmsl@sh\textstyle{#1}{#2}}%
    {\@fmsl@sh\scriptstyle{#1}{#2}}%
    {\@fmsl@sh\scriptscriptstyle{#1}{#2}}}
\def\@fmsl@sh#1#2#3{\m@th\ooalign{$\hfil#1\mkern#2/\hfil$\crcr$#1#3$}}
\makeatother
%\baselineskip 24pt

%%%%%%%%%%%%%%%%%%%%%%%%%%%%%%%%%%%%%%%%%%%%%%%%%%%%%%%%%%%%%%%%%
%%%
%%%                      TITLE PAGE
%%%
%%%%%%%%%%%%%%%%%%%%%%%%%%%%%%%%%%%%%%%%%%%%%%%%%%%%%%%%%%%%%%%%%

\title{\large{\bf Grand Unification on Noncommutative Spacetime}}

\author{Xavier~Calmet\thanks{xcalmet@ulb.ac.be} \\
Service de Physique Th\'eorique, CP225 \\
Boulevard du Triomphe \\
B-1050 Brussels \\
Belgium 
}

\date{June, 2006}

\maketitle

\begin{abstract}
We compute the beta-functions of the standard model formulated on a noncommutative spacetime. If we assume that the scale for spacetime noncommutativity is of the order of  $2.2 \times 10^{15} $ GeV we find that the three gauge couplings of the standard model merge at a scale of $2.3 \times 10^{17}$ GeV. The proton lifetime is thus much longer than in conventional unification models.

\end{abstract}

%%%%%%%%%%%%%%%%%%%%%%%%%%%%%%%%%%%%%%%%%%%%%%%%%%%%%%%%%%%%%%%%
%%%
%%%                     INTRODUCTION
%%%
%%%%%%%%%%%%%%%%%%%%%%%%%%%%%%%%%%%%%%%%%%%%%%%%%%%%%%%%%%%%%%%%

\newpage

%\section{Introduction}

Grand unification \cite{Georgi:1974sy,Fritzsch:1975nn,Georgi:yf} is a topic that has fascinated theoretical physicists since the discovery of the standard model which is based on the gauge symmetry SU(3)$\times$SU(2)$\times$U(1). It is tempting to try to unify these groups within a bigger group such as SU(5) \cite{Georgi:1974sy} or SO(10) \cite{Fritzsch:1975nn}. Unfortunately  the gauge couplings of  the standard model fail to converge to one unified gauge coupling \cite{Amaldi:1991cn} unless one plays with threshold effects \cite{Lavoura:1993su} or breaks the fundamental symmetry of the grand unified gauge group in different steps (see e.g. \cite{Babu:1992ia}). Another way to obtain the unification of the gauge couplings of the standard model is to introduce new particles, e.g.  supersymmetric particles (see e.g. \cite{Dimopoulos:1981yj}), to reach the numerical unification of the gauge couplings.  In this letter we shall pursue a different approach and study whether spacetime noncommutativity can modify the standard model in such a way that the gauge couplings converge to one unified gauge coupling. We do not introduce any new particles and consider a direct breaking of the  grand unified gauge symmetry to the standard model.

Gauge theories formulated on a canonical noncommutative spacetime have
recently received a lot of attention (see e.g. \cite{Seiberg:1999vs,Douglas:2001ba}). A canonical noncommutative
spacetime is defined by the noncommutative algebra
\begin{eqnarray} \label{NCA}
[ \hat x^\mu,\hat x^\nu]=i\theta^{\mu\nu}
\end{eqnarray}
where $\mu$ and $\nu$ run from 0 to 3 and where $\theta^{\mu\nu}$ is
constant and antisymmetric. It has mass dimension minus
two. Formulating Yang-Mills theories relevant to particle physics on
such a spacetime requires one to consider matter fields, gauge fields and
gauge transformations in the enveloping algebra otherwise SU(N) gauge
symmetries cannot be implemented \cite{Madore:2000en,Calmet:2001na}. The enveloping algebra approach allows one to map a noncommutative action $\hat S$ on an effective action formulated on a regular commutative spacetime.

We shall be working within the framework of the minimal noncommutative standard model \cite{Calmet:2001na}, but we nevertheless choose the representation for the U(1) gauge boson in such a way that the trace over three generators of the U(1) is equal to one i.e. instead of taking $Y=diag(-1,1)$ as in  \cite{Calmet:2001na}, we take $Y=1$ which seems to be required in order to have a renormalizable U(1) theory \cite{Calmet:2006zy}. This choice of representation will also help to cure the problem with SU(N) theories which appear at the quantum level \cite{Armoni:2000xr,Bonora:2000ga}. The standard model on a noncommutative spacetime can be written in a very compact way:
\begin{eqnarray} \label{NCSM}
\hat S_{NCSM}&=&\int d^4x \sum_{i=1}^3 \overline{\widehat \Psi}^{(i)}_L \star i
\widehat{\fmslash D} \widehat \Psi^{(i)}_L
+\int d^4x \sum_{i=1}^3 \overline{\widehat \Psi}^{(i)}_R \star i
\widehat{\fmslash  D} \widehat \Psi^{(i)}_R \\
&& \nonumber -\int d^4x \frac{1}{2g'} 
\mbox{{\bf tr}}_{\bf 1} \widehat
F_{\mu \nu} \star  \widehat F^{\mu \nu}
-\int d^4x \frac{1}{2g} \mbox{{\bf tr}}_{\bf 2} \widehat
F_{\mu \nu} \star  \widehat F^{\mu \nu}\\
&&\nonumber
-\int d^4x \frac{1}{2g_3} \mbox{{\bf tr}}_{\bf 3} \widehat
F_{\mu \nu} \star  \widehat F^{\mu \nu}
+ \int d^4x \bigg( \rho_0(\widehat D_\mu \widehat \Phi)^\dagger
\star \rho_0(\widehat D^\mu \widehat \Phi)            
\\ && \nonumber
- \mu^2 \rho_0(\widehat {\Phi})^\dagger \star  \rho_0(\widehat \Phi) - \lambda
\rho_0(\widehat \Phi)^\dagger \star  \rho_0(\widehat \Phi)
\star
\rho_0(\widehat \Phi)^\dagger \star  \rho_0(\widehat \Phi)   \bigg)
\\ && \nonumber
+ \int d^4x \bigg ( 
-\sum_{i,j=1}^3 W^{ij} \bigg
( ( \bar{ \widehat L}^{(i)}_L \star \rho_L(\widehat \Phi))
\star  \widehat e^{(j)}_R
+ \bar {\widehat e}^{(i)}_R \star (\rho_L(\widehat \Phi)^\dagger \star \widehat
L^{(j)}_L) \bigg )
\\ && \nonumber
-\sum_{i,j=1}^3 G_u^{ij} \bigg
( ( \bar{\widehat Q}^{(i)}_L \star \rho_{\bar Q}(\widehat{\bar\Phi}))\star  
\widehat u^{(j)}_R
+ \bar {\widehat u}^{(i)}_R \star 
(\rho_{\bar Q}(\widehat{\bar\Phi})^\dagger
\star \widehat Q^{(j)}_L) \bigg )
\\ && \nonumber
-\sum_{i,j=1}^3 G_d^{ij} \bigg
( ( \bar{ \widehat Q}^{(i)}_L \star \rho_Q(\widehat \Phi))\star  
\widehat d^{(j)}_R
+ \bar{ \widehat d}^{(i)}_R \star (\rho_Q(\widehat \Phi)^\dagger
\star \widehat Q^{(j)}_L) \bigg ) \bigg),
\end{eqnarray}
where $\star$ is the star product,  $\bar{\Phi} = i \tau_2 \Phi^*$ and $\rho(\hat F)$ denotes the representation in the enveloping algebra of the field $\hat F$ (see  \cite{Calmet:2001na} for details).
The matrices $W^{ij}$, $G^{ij}_u$ and $G^{ij}_d$ are
the Yukawa couplings. Note that in (\ref{NCSM}) we have not yet developed the fields in the enveloping algebra.  The action (\ref{NCSM}) has the standard model as a limit for $\theta\to 0$:
\begin{eqnarray} \label{SM}
\hat S_{NCSM}&=&\int d^4x \sum_{i=1}^3 \overline{\Psi}^{(i)}_L  i
{\fmslash D}  \Psi^{(i)}_L
+\int d^4x \sum_{i=1}^3 \overline{ \Psi}^{(i)}_R  i
{\fmslash  D} \Psi^{(i)}_R \\
&& \nonumber 
-\int d^4x \frac{1}{4} 
f_{\mu \nu} f^{\mu \nu}
-\int d^4x \frac{1}{2} \mbox{Tr}
F_{\mu \nu}  F^{\mu \nu}
-\int d^4x \frac{1}{2} \mbox{Tr}
G_{\mu \nu} G^{\mu \nu} \\
\nonumber &&
+ \int d^4x \bigg( (D_\mu \Phi)^\dagger (D^\mu  \Phi)            
- \mu^2 \Phi^\dagger  \Phi- \lambda (\Phi^\dagger \Phi)^2  \bigg)
\\ && \nonumber
+ \int d^4x \bigg ( 
-\sum_{i,j=1}^3 W^{ij} \bigg (  \bar{L}^{(i)}_L  \Phi e^{(j)}_R
+ \bar {e}^{(i)}_R \Phi^\dagger L^{(j)}_L \bigg )
\\ && \nonumber
-\sum_{i,j=1}^3 G_u^{ij} \bigg (
\bar{Q}^{(i)}_L \bar\Phi
 u^{(j)}_R
+ \bar {u}^{(i)}_R 
\bar\Phi^\dagger Q^{(j)}_L) \bigg )
\\ && \nonumber
-\sum_{i,j=1}^3 G_d^{ij} \bigg
(  \bar{ Q}^{(i)}_L \Phi  d^{(j)}_R
+ \bar{ d}^{(i)}_R  \Phi^\dagger Q^{(j)}_L \bigg ) \bigg)+ {\cal O}(\theta).
\end{eqnarray}

Using the quantization and regularization methods presented in \cite{Calmet:2006zy} we now compute the $\beta$-functions of this model at the one loop approximation. The Feynman rules are given by
\begin{eqnarray}
 \hat \Psi(p^I)  \hat A^\mu \hat \Psi(p^F) &\to& i g \gamma^\mu \exp(
 \frac{i}{2} p_\alpha^I \theta^{\alpha\beta} p^F_\beta) \rho(\hat A^\mu)
\end{eqnarray}
where $\rho(\hat A^\mu)$ denotes the representation of the gauge boson $\hat A^\mu$ in the enveloping algebra, note that $\rho( A^\mu)=T^a$ for the commutative field. The Feynman rule for the three gauge boson interaction is given by
 \begin{eqnarray}
 \hat A^{\mu_1}(p^1)  \hat B^{\mu_2}(p^2)\hat C^{\mu_3}(p^3) &\to& - g  (f_{ABC} \cos(p^1_\alpha \theta^{\alpha\beta} p^2_\beta) + d_{ABC} \sin(p^1_\alpha \theta^{\alpha\beta} p^2_\beta ))
  \\ && \nonumber
 \times [(p^1-p^2)^{\mu_3} g^{\mu_1\mu_2} + (p^2-p^3)^{\mu_1} g^{\mu_2\mu_3}
 +(p^3-p^1)^{\mu_2} g^{\mu_3\mu_1}] ,
 \end{eqnarray}
 where $f_{ABC}=-2 i Tr(
 [ \rho(\hat A^{\mu_1}(p^1)), \rho(\hat B^{\mu_2}(p^2))]  \rho(\hat C^{\mu_2}(p^3)))$ is traced in the enveloping algebra as well as $d_{ABC}=2 Tr( \rho(
 \{\hat A^{\mu_1}(p^1)), \rho(\hat B^{\mu_2}(p^2))\}  \rho(\hat C^{\mu_2}(p^3)))$, i.e. they should not be confused with the usual group theoretical factors  $i f_{abc} t^c=[t_a,t_b]$ and $ d_{abc} t^c=\{t_a,t_b\}$ although we shall see that they are related to these factors once one has expanded the trace in the enveloping algebra. For the four bosons interaction, one has
 \begin{eqnarray}
 \hat A^{\mu_1}(p^1) \hat
 B^{\mu_2}(p^2) \hat C^{\mu_3}(p^3) \hat D^{\mu_4}(p^4) &\to& 
 -  i \sum_x g^2 [ 
(f_{ABx} \cos(p^1_\alpha \theta^{\alpha\beta} p^2_\beta) + d_{ABx} \sin(p^1_\alpha \theta^{\alpha\beta} p^2_\beta ))
\\ \nonumber && \times
(f_{xCD} \cos(p^3_\alpha \theta^{\alpha\beta} p^4_\beta) + d_{xCD} \sin(p^3_\alpha \theta^{\alpha\beta} p^4_\beta )) \\ \nonumber && \times
( g^{\mu_1\mu_3}g^{\mu_2\mu_4} - g^{\mu_1\mu_4}g^{\mu_2\mu_3})  \\ && \nonumber
+(f_{ACx} \cos(p^1_\alpha \theta^{\alpha\beta} p^3_\beta) + d_{ACx} \sin(p^1_\alpha \theta^{\alpha\beta} p^3_\beta ))
\\ \nonumber &&
(f_{xDB} \cos(p^4_\alpha \theta^{\alpha\beta} p^2_\beta) + d_{xDB} \sin(p^4_\alpha \theta^{\alpha\beta} p^2_\beta )) \\ \nonumber && \times
  ( g^{\mu_1\mu_4}g^{\mu_2\mu_3} -
 g^{\mu_1\mu_2}g^{\mu_3\mu_4}) 
 \\ \nonumber && 
  +
(f_{ADx} \cos(p^1_\alpha \theta^{\alpha\beta} p^4_\beta) + d_{ADx} \sin(p^1_\alpha \theta^{\alpha\beta} p^4_\beta )) \\ \nonumber && \times
(f_{xBC} \cos(p^2_\alpha \theta^{\alpha\beta} p^3_\beta) + d_{xBC} \sin(p^2_\alpha \theta^{\alpha\beta} p^3_\beta )) \\ \nonumber && \times
 ( g^{\mu_1\mu_2}g^{\mu_3\mu_4} - g^{\mu_1\mu_3}g^{\mu_2\mu_4})] 
 \end{eqnarray}
and the interaction for the ghost field is given by
\begin{eqnarray}
 \hat G_1(p^I) \hat A^\mu(q)
 \hat G_2(p^F) &\to& - g p^F_\mu (\cos(  p_\alpha^I
 \theta^{\alpha\beta} p^F_\beta) f_{AG_2G_1}-\sin(  p_\alpha^I
 \theta^{\alpha\beta} p^F_\beta) d_{AG_2G_1} ). 
\end{eqnarray}
The gauge boson $\hat A_\mu$ is valued in the enveloping algebra and contains all the gauge bosons of the standard model. The first order expansion in the enveloping of  $\hat A_\mu= g' Y {\cal A}_\mu + g B_\mu^a \tau^a + g_s G_\mu^c T^c$.
Our aim is not to study the renormalizability of the model, but only to extract the UV divergent part  necessary for the calculation of the beta-functions. Let us start with the gauge boson  $\hat A_\mu$.  As usual there are three self-energy diagrams of interest. It is easy to see that the planar diagram contribution to the UV divergent part  is given by  
\begin{eqnarray}
D= g^2 \left ( 
 2 Tr( \{\rho(\hat E^{\mu_1}(p^3)), \rho(\hat F^{\mu_2}(p^4))\}  \rho(\hat A^{\mu_2}(p^1))) \cos(\frac{1}{2} p^1_\alpha \theta^{\alpha\beta} p^2_\beta) \right.  \\ \left.  \nonumber 
 +  Tr( [\rho( \hat E^{\mu_1}(p^3)), \rho(\hat F^{\mu_2}(p^4))]  \rho(\hat A^{\mu_2}(p^1))) \sin(\frac{1}{2} p^1_\alpha \theta^{\alpha\beta} p^2_\beta) \right) 
 \\ \nonumber \times 
 \left ( 
 2 Tr( \{\rho(\hat E^{\mu_1}(p^3)), \rho(\hat F^{\mu_2}(p^4))\}  \rho(\hat B^{\mu_2}(p^2))) \cos(\frac{1}{2} p^1_\alpha \theta^{\alpha\beta} p^2_\beta) \right.  \\ \left.  \nonumber 
 +  Tr( [\rho( \hat E^{\mu_1}(p^3)), \rho(\hat F^{\mu_2}(p^4))]  \rho(\hat B^{\mu_2}(p^2))) \sin(\frac{1}{2} p^1_\alpha \theta^{\alpha\beta} p^2_\beta) \right). 
 \end{eqnarray}
 We are interested in the contribution to the running of the gauge fields after they have been mapped on a commutative spacetime as we want to compare the running of the three gauge couplings to that of the regular standard model. Let us first consider the case where the outer particles are SU(N) gauge bosons as well as the inner one. One finds 
 \begin{eqnarray}
 2 Tr( \{\rho(\hat E^{\mu_1}(p^3)), \rho(\hat F^{\mu_2}(p^4))\}  \rho(\hat A^{\mu_2}(p^1)))= -2 iTr([T^e,T^f]T^a)= f^{efa}
\end{eqnarray}
   and  
  \begin{eqnarray}
 2 Tr( [\rho(\hat E^{\mu_1}(p^3)), \rho(\hat F^{\mu_2}(p^4))]  \rho(\hat A^{\mu_2}(p^1)))= 2 Tr(\{T^e,T^f\}T^a)= d^{efa}
  \end{eqnarray}
    in the leading order of the expansion in the enveloping algebra and one thus finds 
 \begin{eqnarray}
D^{(1)}_{SU(N)}= g^2 N \left ( 1- \frac{1}{N} \sin^2((\frac{1}{2} p^1_\alpha \theta^{\alpha\beta} p^2_\beta) \right) 
 \end{eqnarray}
which corresponds to the result obtained  in refs. \cite{Armoni:2000xr,Bonora:2000ga}. Now in the noncommutative standard model there is a U(1) piece that couples to the SU(N) bosons, since we have chosen $Y=1$, calculating the trace in the enveloping algebra is easy 
 \begin{eqnarray}
2 Tr( [\rho(\hat E^{\mu_1}(p^3)), \rho(\hat F^{\mu_2}(p^4))]  \rho(\hat A^{\mu_2}(p^1))= -2 iTr([T^e,Y]T^a)= 0
\end{eqnarray}
and 
 \begin{eqnarray}
2 Tr( \{\rho(\hat E^{\mu_1}(p^3)), \rho(\hat F^{\mu_2}(p^4))\}  \rho(\hat A^{\mu_2}(p^1))=2Tr(\{T^e,Y\}T^a)=2/N
\end{eqnarray}
 and the second contribution is thus given by
 \begin{eqnarray}
D^{(2)}_{SU(N)}=2 g^2 \frac{2}{N}  \sin^2(\frac{1}{2} p^1_\alpha \theta^{\alpha\beta} p^2_\beta).
 \end{eqnarray}
The total contribution is thus given by $g^2N$ and is identical to that of the standard model. It is easy to see that the contribution to the divergence in the U(1) sector is that given in \cite{Calmet:2006zy}. However there are two new contributions coming from the SU(2) and SU(3) sectors and the divergent part of the self-interaction diagrams is given by $5/3(1+2+3)$. 

We have seen that the contribution of the SU(2) and SU(3) gauge bosons are thus, for this choice of representation in the enveloping algebra for the U(1) sector, the same as in the standard model on a commutative spacetime. Whereas it is clear, that a pure SU(N) noncommutative gauge theory is not renormalizable as shown in  \cite{Armoni:2000xr,Bonora:2000ga}, the standard model on a noncommutative spacetime based on the enveloping algebra of su(3)$\times$su(2)$\times$u(1) has a chance to be renormalizable due to the freedom in the choice of the representations of the U(1) sector. We leave that question open. 
The contribution from the fermion and the scalar field can be deduced from the quantum electrodynamics contribution studied in \cite{Calmet:2006zy}. We thus find:
\begin{eqnarray}
\frac{\partial}{\partial \mu} \alpha_i(\mu) = \frac{1}{2 \pi} b^{NC}_i \alpha^2_i(\mu), \ i\in \{1,2,3\}
\end{eqnarray}
with 
\begin{eqnarray}
b^{NC}_i =\left(
\begin{array}{c}
    b^{NC}_1 \\
   b^{NC}_2  \\
 b^{NC}_3   
\end{array}
\right)
=
\left(
\begin{array}{c}
    -12 \\
   -22/3  \\
 -11  
\end{array}
\right)+ N_{f} 
\left(
\begin{array}{c}
    4/3 \\
   4/3  \\
 4/3  
\end{array}
\right)
+N_{Higgs}
\left(
\begin{array}{c}
    1/10 \\
   1/6  \\
 0  
\end{array}
\right),
\end{eqnarray}
where $\alpha_i=$ where $g_1$ is the U(1) gauge coupling, $g_2$ the SU(2) one , $g_3$ the SU(3) one, $N_f=3$ is the number of families and $N_{Higgs}=1$ is the number of Higgs bosons. If we compare the $b_i$ of the standard model on a noncommutative spacetime to that of the standard model on a commutative spacetime
\begin{eqnarray}
b^{SM}_i =\left(
\begin{array}{c}
    b^{SM}_1 \\
   b^{SM}_2  \\
 b^{SM}_3   
\end{array}
\right)
=
\left(
\begin{array}{c}
    0 \\
   -22/3  \\
 -11  
\end{array}
\right)+ N_{f} 
\left(
\begin{array}{c}
    4/3 \\
   4/3  \\
 4/3  
\end{array}
\right)
+N_{Higgs}
\left(
\begin{array}{c}
    1/10 \\
   1/6  \\
 0  
\end{array}
\right),
\end{eqnarray}
we see that the only difference is the factor $-12$ for the U(1) gauge coupling which comes from the nonabelian like term in the U(1)  noncommutative gauge boson interaction.

We shall now study the grand unification of the gauge couplings. Clearly the running of the beta-function of the  U(1) sector is in contradiction with experiment. The noncommutative parameter $\theta$ thus has to be space-time dependent or in other words energy-momentum dependent. It has been shown how to formulate Yang-Mills theories on a space-time with an energy-momentum dependent noncommutativity \cite{Calmet:2003jv}.  We shall not go into these details and treat the scale dependence of $\theta$ as a threshold effect. We shall assume that $\theta^{\mu\nu}=0$ for $\mu < \Lambda_{NC}$ and  $\theta^{\mu\nu}\neq 0$ for  $\mu \ge\Lambda_{NC}$, i.e. the noncommutativity of spacetime can only be probed when one goes to short enough distances. We thus have one free parameter $\Lambda_{NC}$. We know from experimental bounds that $\Lambda_{NC}>{\cal O}$(1 TeV) \cite{Calmet:2004dn}, however if spacetime noncommutativity is responsible for the unification of the gauge couplings of the standard model we will see that the typical scale for spacetime noncommutativity is much higher and is out of reach of future colliders.

Taking the following input values \cite{Amaldi:1991cn} $\alpha_1(M_Z)=0.0168$,  $\alpha_2(M_Z)=0.03322$ and $\alpha_3(M_Z)=0.118$, we find that if we assume that the scale for spacetime noncommutativity is $2.2 \times 10^{15} $ GeV, the three gauge couplings of the standard model unify at a scale of $\Lambda_u=2.3 \times 10^{17}$ GeV and the unified gauge coupling $\alpha_u$ is equal to 0.0208 at the unification scale. Grand unification within the noncommutative setting avoids problems with the proton decay as the grand unification scale is much higher than in conventional unification models. Unfortunately, it seems hopeless to test such a long proton lifetime. As in any non-supersymmetric unified theory, e.g. SU(5), we expect the nucleon lifetime to be given by a dimension 6 operator which is suppressed by  the unification scale squared i.e. the proton lifetime is given by:
\begin{eqnarray} \label{plife}
\tau_p \propto \frac{\Lambda_u^4}{\alpha_u^2 m_p^5}=1.8 \times 10^{41} \mbox{yr},
\end{eqnarray}
where $m_p$ is the proton mass. The present limit \cite{Eidelman:2004wy} on the proton lifetime is of the order of $10^{33}$ yr (this limit is obviously decay channel dependent).  The result  (\ref{plife}) is clearly out of reach for present or future experiments.  However, noncommutative grand unification \cite{He:2002mz} is  a viable alternative to supersymmetric unification and does not involve any new particle. It is also interesting to note that the scale $2.3 \times 10^{17}$ GeV is not very far away from the Planck scale. 

One of the main motivations to consider a noncommutative spacetime is that it introduces the notion of minimal length in quantum field theoretical models. A minimal length is a natural expectation of a unified theory of quantum mechanics and general relativity \cite{minlength1}. The effects of such a minimal length are expected to become relevant close to the Planck scale, which corresponds to an energy scale of the order of $10^{19}$ GeV, scale at which gravity would unify with gauge theories as in e.g. string theory. It is very interesting to note that the scale for the gauge unification on a noncommutative spacetime is quite close to the scale where one expects gravity to unify with the other forces of nature

\bigskip
%%%%%%%%%%%%%%%%%%%%%%%%%%%%%%%%%%%%%%%%%%%%%%%%%%%%%%%%%%%%%%%%%
%%%
%%%                   ACKNOWLEDGMENTS
%%%
%%%%%%%%%%%%%%%%%%%%%%%%%%%%%%%%%%%%%%%%%%%%%%%%%%%%%%%%%%%%%%%%%
\subsection*{Acknowledgments}
\noindent 
The author would like to thank Jean-Marie Fr\`ere for helpful discussions. This work was supported in part by the IISN and the Belgian science
policy office (IAP V/27).
%%%%%%%%%%%%%%%%%%%%%%%%%%%%%%%%%%%%%%%%%%%%%%%%%%%%%%%%%%%%%%%%%
%%%
%%%                     BIBLIOGRAPHY
%%%%%%%%%%%%%%%%%%%%%%%%%%%%%%%%%%%%%%%%%%%%%%%%%%%%%%%%%%%%%%%%%%%%

\bigskip

%\newpage
%\vskip .75 in

\end{document}